\documentclass[prb,aps,twocolumn]{revtex4-1}

\usepackage{amsmath,amssymb,amstext}

\usepackage{txfonts}

\usepackage{lmodern}

\usepackage{graphicx}

\usepackage[hang, nooneline]{subfigure}

\newcommand{\vect}[1]{\boldsymbol{#1}}
\newcommand{\im}{\mathrm i \,}
\newcommand{\e}[1]{\, \mathrm e^{\, #1 }}

\begin{document}

\title{Intrinsic Spin Swapping}
\author{Severin Sadjina and Arne Brataas}
\affiliation{Department of Physics, Norwegian University of Science and Technology,
NO-7491 Trondheim, Norway}
\author{A.~G. Mal'shukov}
\affiliation{Institute of Spectroscopy, Russian Academy of
Sciences, 142190, Troitsk, Moscow oblast, Russia}

\begin{abstract}
Here, we study diffusive spin transport in two dimensions and demonstrate that an intrinsic analog to a previously predicted extrinsic spin swapping effect, where the spin polarization and the direction of flow are interchanged due to spin-orbit coupling at extrinsic impurities, can be induced by intrinsic (Rashba) spin-orbit coupling. The resulting accumulation of intrinsically spin-swapped polarizations is shown to be much larger than for the extrinsic effect. Intrinsic spin swapping is particularly strong when the system dimensions exceed the spin-orbit precession length and the generated transverse spin currents are of the order of the injected primary spin currents. In contrast, spin accumulations and spin currents caused by extrinsic spin swapping are proportional to the spin-orbit coupling. We present numerical and analytical results for the secondary spin currents and accumulations generated by intrinsic spin swapping, and we derive analytic expressions for the induced spin accumulation at the edges of a narrow strip, where a long-range propagation of spin polarizations takes place.
\end{abstract}

\maketitle

%------------------------------------------------------------
% Introduction
%------------------------------------------------------------

   \section{Introduction}
   \label{sec:introduction}

   Understanding the spin-orbit interaction is essential to the development of spintronics and gives rise to various spin transport mechanisms. Effects of the spin-orbit interaction can be intrinsic or extrinsic. Intrinsic effects are caused by the spin-orbit interaction in the band structure. Extrinsic contributions arise from spin-orbit coupling at impurities. The spin Hall effect, where a transverse spin current is generated via a longitudinal charge current, is one of the effects resulting from the spin-orbit coupling and has attracted much attention, both theoretically\cite{Karplus1954, Dyakonov1971b,Dyakonov1971,Hirsch1999,Zhang2000,Murakami2003,Sinova2004,Engel2007} and experimentally\cite{Kato2004,Sih2005,Wunderlich2005,Saitoh2006,Stern2006,Valenzuela2006,Kimura2007}. In a sample, this transverse spin current generates opposite spin accumulations at the lateral boundaries.
   
   While the spin Hall effect provides coupling between charge and spin, another spin-orbit-induced transport mechanism has recently been introduced in which only spins couple and which emerges even in the absence of charge currents. Primary longitudinal spin currents give rise to secondary transverse spin currents due to spin-orbit coupling at extrinsic impurities.\cite{Lifshits2009} The generated secondary spin currents are proportional to the extrinsic spin-orbit coupling strength. The effect has been coined `spin swapping' because, in its simplest manifestation, it interchanges the spin polarization direction and the spin flow.\footnote{The terms responsible for extrinsic spin swapping were already indicated in Refs.\ \protect\onlinecite{Dyakonov1971b,Dyakonov1971}. However, their physical origin was not understood at the time\protect\cite{Lifshits2009}} It has been suggested that any mechanism inducing a spin Hall effect should also give rise to spin swapping. However, it has not yet been clear how the intrinsic mechanism could produce this effect.

   In this paper, we demonstrate that an intrinsic (Rash\-ba spin-orbit-induced) spin swapping effect exists in two-dimensional diffusive metals and that it is drastically different from its extrinsic analog. The main distinction between these two effects is that the extrinsic effect is of the same order as the spin-orbit coupling strength and is thus small, irrespective of the system size. In contrast, the intrinsic spin swapping effect is large for system dimensions exceeding the spin-orbit precession length, and the secondary spin currents generated by this effect are then of the same order as the primary spin currents. If, however, the system width is small compared to the spin-orbit precession length, the effect is small but leads to a long-range propagation of spin polarizations closely related to the increase of the D'yakonov-Perel spin relaxation time in narrow strips\cite{Malsh2000}. Furthermore, the symmetry of intrinsic spin swapping is more complex and richer than that of the extrinsic spin swapping effect resulting in a non-trivial dependence on the relative orientation of the injected spin flow and the spin polarization. We present numerical and analytical results for the transverse secondary spin currents and accumulations induced by primary spin currents in two-dimensional diffusive metals, and we compare the intrinsic and extrinsic spin swapping effects.
   
   This paper is organized as follows. We first provide a review of the previously discussed extrinsic spin swapping effect in Sec.\ \ref{sec:extrinsic}, and we compute the spin accumulations and spin currents induced by an injected primary spin current in a two-dimensional diffusive metal. In Sec.\ \ref{sec:intrinsic}, we discuss the intrinsic spin swapping effect, numerically evaluate the spin densities and spin currents generated through intrinsic spin swapping, and derive analytical results for the resulting spin currents far away from the lateral edges of a sample. Next, in Sec.\ \ref{sec:narrowstrip}, we treat the case of a narrow strip whose width is small compared to the spin-orbit precession length and find analytical expressions for the spin accumulations at the lateral edges of a sample stemming from the intrinsic spin swapping effect. In Sec.\ \ref{sec:observation}, we briefly discuss how the spin swapping effects could be observed in experiment. Finally, we give our conclusions in Sec.\ \ref{sec:conclusion}.

%------------------------------------------------------------
% Extrinsic spin swapping
%------------------------------------------------------------

   \section{Extrinsic spin swapping}
   \label{sec:extrinsic}

   First, we review the extrinsic spin swapping effect introduced in Ref.\ \onlinecite{Lifshits2009} and present its features in two-dimensional diffusive metals in order to compare it to the intrinsic spin swapping effect to be discussed later. The Hamiltonian of the system under consideration reads
    % -------------------------------------------------------
    % Hamiltonian
    \begin{equation}
    \label{eq:extrinsic_hamiltonian}
      \uppercase{\mathcal{H}}(\vect{\rho})
      =
      - \frac{1}{2m} \partial^{2}_{\vect{\rho}}
      +
      V_\text{imp}(\vect{\rho})
      +
      V_\text{so}(\vect{\rho}),
    \end{equation}
    % - - - - - - - - - - - - - - - - - - - - - - - - - - - -
    where $\vect{\rho}=(x,y)$ is a two-dimensional coordinate,
    % -------------------------------------------------------
    % Elastic Impurity Scattering
    \begin{subequations}
    \label{eq:impurity}
    \begin{equation}
    \label{eq:impurity_simple}
      V_\text{imp}(\vect{\rho})
      =
      \frac{1}{A} \sum_{\vect{\rho}_i} \sum_{\vect{k}} v(\vect{k}) \e{\im \vect{k} \cdot (\vect{\rho} - \vect{\rho}_i)}
    \end{equation}
    % - - - - - - - - - - - - - - - - - - - - - - - - - - - -
    is the elastic impurity scattering potential, and
    % -------------------------------------------------------
    % Elements of U
    \begin{equation}
    \label{eq:impurity_soi}
      V_\text{so}(\vect{\rho})
      =
      - \im \gamma \Big[ \vect{\sigma} \times \vect{\nabla} V_\text{imp}(\vect{\rho}) \Big] \cdot \vect{\partial}_{\vect{\rho}}
    \end{equation}
    \end{subequations}
    % - - - - - - - - - - - - - - - - - - - - - - - - - - - -
    is the spin-orbit coupling. $\vect{\rho}_i$ is the position of the $i^{\text{th}}$ impurity, $A$ is the area, $v(\vect{k})$ is the Fourier transformed scattering potential, $\vect{\sigma} = (\sigma_x, \sigma_y, \sigma_z)^\text{T}$ is a vector of Pauli matrices, and $\gamma$ is the dimensionless spin-orbit coupling strength.
    
    Considering transport in the diffusive limit, the spin diffusion equation reads
    % -------------------------------------------------------
    % Spin diffusion
    \begin{equation}
    \label{eq:extrinsic_spin_diffusion}
      \vect{\partial}^{2}_{\vect{\rho}} f_b
      -
      \frac{1}{l_\text{sf}^2}
      f_b
      =0
      ,
    \end{equation}
    % - - - - - - - - - - - - - - - - - - - - - - - - - - - -
    where $f_b$ is the $b$ component of the spin density, $b \in \{x,y,z\}$, and $l_\text{sf}$ is the spin-flip length. In order to study spin transport, one also needs to define the spin current. In the leading approximation, while neglecting spin-orbit effects, the spin current is given by the spin diffusion current  $j_{a b}^{(0)} = - D \partial_a f_b$ flowing along $a$ and polarized along $b$, where $D$ is the diffusion constant. The spin-orbit interaction gives rise to additional terms in the spin current. To first order in the spin-orbit coupling strength $\gamma$, when there is no charge current giving rise to the spin Hall effect, the spin current is\cite{Lifshits2009}
    % -------------------------------------------------------
    % Spin Current
    \begin{equation}
    \label{eq:normal_spin_current}
      j_{a b}
      =
      j_{a b}^{(0)}
      +
      \chi
      \Big(
        j_{b a}^{(0)}
	-
	\delta_{a b}
	j_{c c}^{(0)}
      \Big)
      .
    \end{equation}
    % - - - - - - - - - - - - - - - - - - - - - - - - - - - -
    The term proportional to the swapping constant $\chi$ relates the spin polarization to the direction of flow and results in the induction of secondary spin currents, i.e., a `spin swapping' effect.\cite{Lifshits2009} For example, a primary spin current directed along $x$ will induce transverse spin currents that arise as follows,
    % -------------------------------------------------------
    % Induced transverse Spin Currents
    \begin{align*}
      j_{x b}^{(0)} &\Rightarrow j_{b x},\\
    \intertext{if $b \neq x$, and}
      j_{x x}^{(0)} &\Rightarrow - j_{y y} - j_{z z}.
    \end{align*}
    % - - - - - - - - - - - - - - - - - - - - - - - - - - - -
    The first of these transformations swaps the current's flow direction and its polarization. In general, this causes spin accumulations at the lateral edges of a sample, as we shall see shortly. The swapping constant is linear in the spin-orbit coupling strength and can be calculated explicitly, \cite{Lifshits2009}
    % -------------------------------------------------------
    % Spin Current
    \begin{equation}
    \label{eq:normal_swappingconstant}
      \chi = 2 \gamma p_\text{F}^2
      ,
    \end{equation}
    % - - - - - - - - - - - - - - - - - - - - - - - - - - - -
    where $p_\text{F}$ is the momentum at the Fermi level and short-ranged scattering potentials are assumed.
    
    Extrinsic spin swapping arises from the additional terms in the spin current \eqref{eq:normal_spin_current} that are proportional to $\chi$, whereas the spin diffusion equation \eqref{eq:extrinsic_spin_diffusion} is unaltered. Extrinsic spin swapping therefore affects the boundary conditions for an unaltered, conventional spin diffusion differential equation. We will see later that the spin diffusion equation for the intrinsic spin swapping effect is altered as well, giving rise to a richer class of phenomena.
    \begin{figure}[h!tb]
    \centering
      \subfigure[Spin density and spin current polarized along $x$]
      {\includegraphics[scale=0.35]{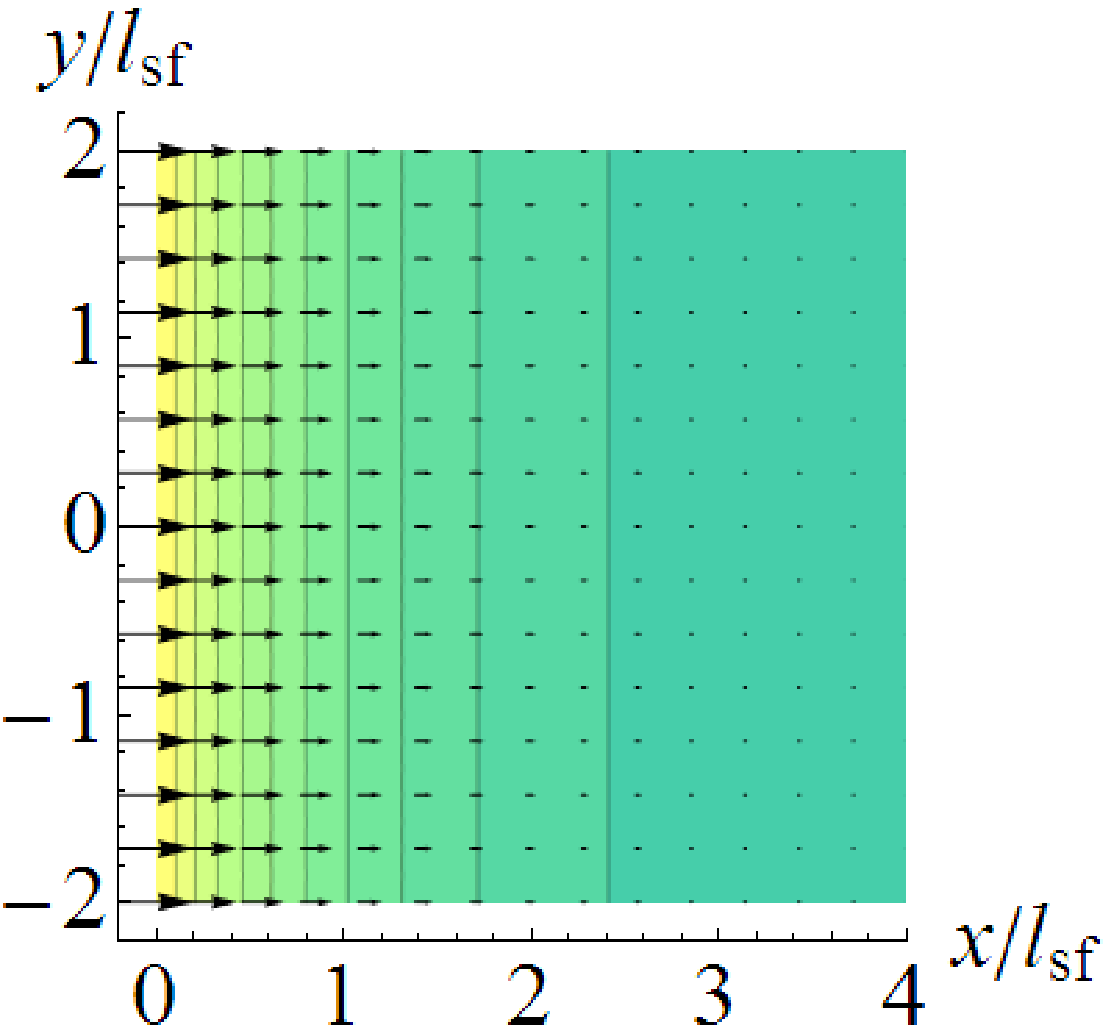}
      \label{fig:ext_fx}
	}
      \subfigure[Spin density and spin current polarized along $y$]{
      \includegraphics[scale=0.35]{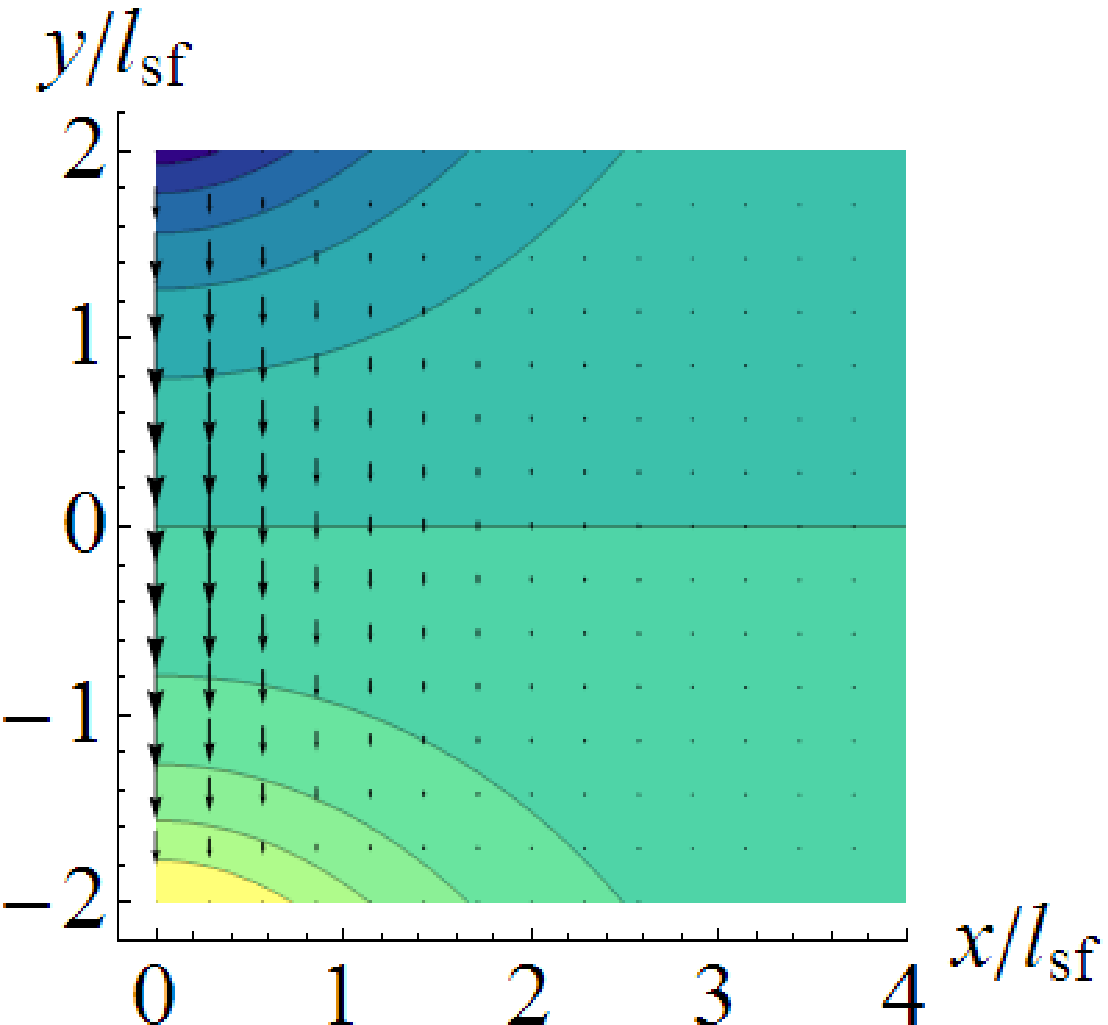}
      \label{fig:ext_fy}
	}\\
      \subfigure{
	\hspace{0pt}\setlength\fboxsep{2pt}
\setlength\fboxrule{0.25pt}
      \fbox{\includegraphics[scale=0.46]{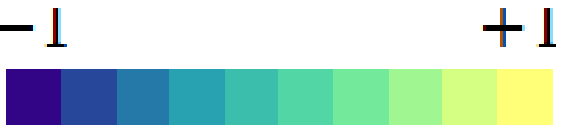}}
	}\\
      \caption{(Color online) The extrinsic spin swapping effect in a semi-infinite two-dimensional diffusive metal of width $L = 4 l_\text{sf}$. Shown are the scaled spin densities and the scaled spin currents according to Eqs.\ \eqref{eq:extrinsic_spin_diffusion} and \eqref{eq:normal_spin_current}. A primary spin current $j_{x x}^{(0)}$ injected at $x=0$ in \subref{fig:ext_fx} induces a transverse spin current $j_{y y}$ through the spin swapping effect in \subref{fig:ext_fy}. Note that the secondary spin accumulation and spin currents in \subref{fig:ext_fy} are linear with respect to the small swapping constant $\chi$. If, instead, a primary spin current $j_{x y}^{(0)}$ polarized along $y$ is injected, $f_y$ and $j_{x y}$ are illustrated by \subref{fig:ext_fx} and the resulting secondary spin density and spin current ($f_x$ and $j_{y x}$, respectively) only differ from \subref{fig:ext_fy} by a sign.}
    \label{fig:ext_dens}
    \end{figure}

    In order to compare the extrinsic spin swapping effect with its intrinsic analog to be discussed in the next section, we first study the spin polarizations generated via the extrinsic spin swapping effect beyond the discussion given in Ref.\ \onlinecite{Lifshits2009}. We consider a semi-infinite two-dimensional diffusive metal of width $L$ into which a spin current $j_{x x}^{(0)}$ directed along $x$ and carrying spins polarized along $x$ is injected at $x = 0$. We assume that the injected current is homogeneous along $y$ at the injection edge. Further, we assume impenetrable lateral sample edges such that no spin current flows through, with $j_{y b}(y = \pm L/2) = 0$ for any spin polarization $b$. The spin-orbit coupling at extrinsic impurities generates a transverse spin current $j_{y y}$ on length scales larger than the mean free path according to Eq.\ \eqref{eq:normal_spin_current}. In turn, this gives rise to an accumulation of spins at the lateral edges of the sample polarized along $y$ that is anti-symmetric in the transverse coordinate $y$. The spin accumulation and spin current are plotted in Fig.\ \ref{fig:ext_dens}: Fig.\ \ref{fig:ext_fx} shows the polarization along $x$, and Fig.\ \ref{fig:ext_fy} shows the polarization along $y$. In the two-dimensional case considered here, no transformation into spins polarized along $z$ takes place. Note that the extrinsic spin swapping effect and, therefore, the resulting secondary spin accumulations and spin currents are of the order of the small swapping constant $\chi$. Solving the spin diffusion equation with the above-mentioned boundary conditions, the accumulation of spins at the lateral edges of a sample can be obtained analytically and may be probed experimentally,
    % -------------------------------------------------------
    % Lateral Spin Accumulation
    \begin{equation*}
      f_y ( y = \pm L/2 )
      =
      \left\{
      \begin{aligned}
      &\pm \frac{2}{\pi} \frac{j_{x x}^{(0)} \chi}{D} x K_1 (x/l_\text{sf}) ,& \text{for} \quad &L \gg l_\text{sf},
      \\
      &\pm \frac{L}{2} \frac{j_{x x}^{(0)} \chi}{D} \e{-x/l_\text{sf}},& \text{for} \quad &L \ll l_\text{sf},
      \end{aligned}
      \right.
    \end{equation*}
    % - - - - - - - - - - - - - - - - - - - - - - - - - - - -
    where $K_1$ is the modified Bessel function of the second kind and first order. This coincides with the numerical result illustrated in Fig.\ \ref{fig:ext_dens}. If, instead, a primary spin current $j_{x y}^{(0)}$ polarized along $y$ is injected, the resulting secondary spin densities and spin currents ($f_x$ and $j_{y x}$, respectively) differ only by a sign according to Eq.\ \eqref{eq:normal_spin_current} and can also be illustrated as shown in Fig.\ \ref{fig:ext_dens}.

    \begin{widetext}
    
    \begin{figure}[h!tb]
    \centering
    \subfigure[Spin density and spin current polarized along $x$]{
      \includegraphics[scale=0.46]{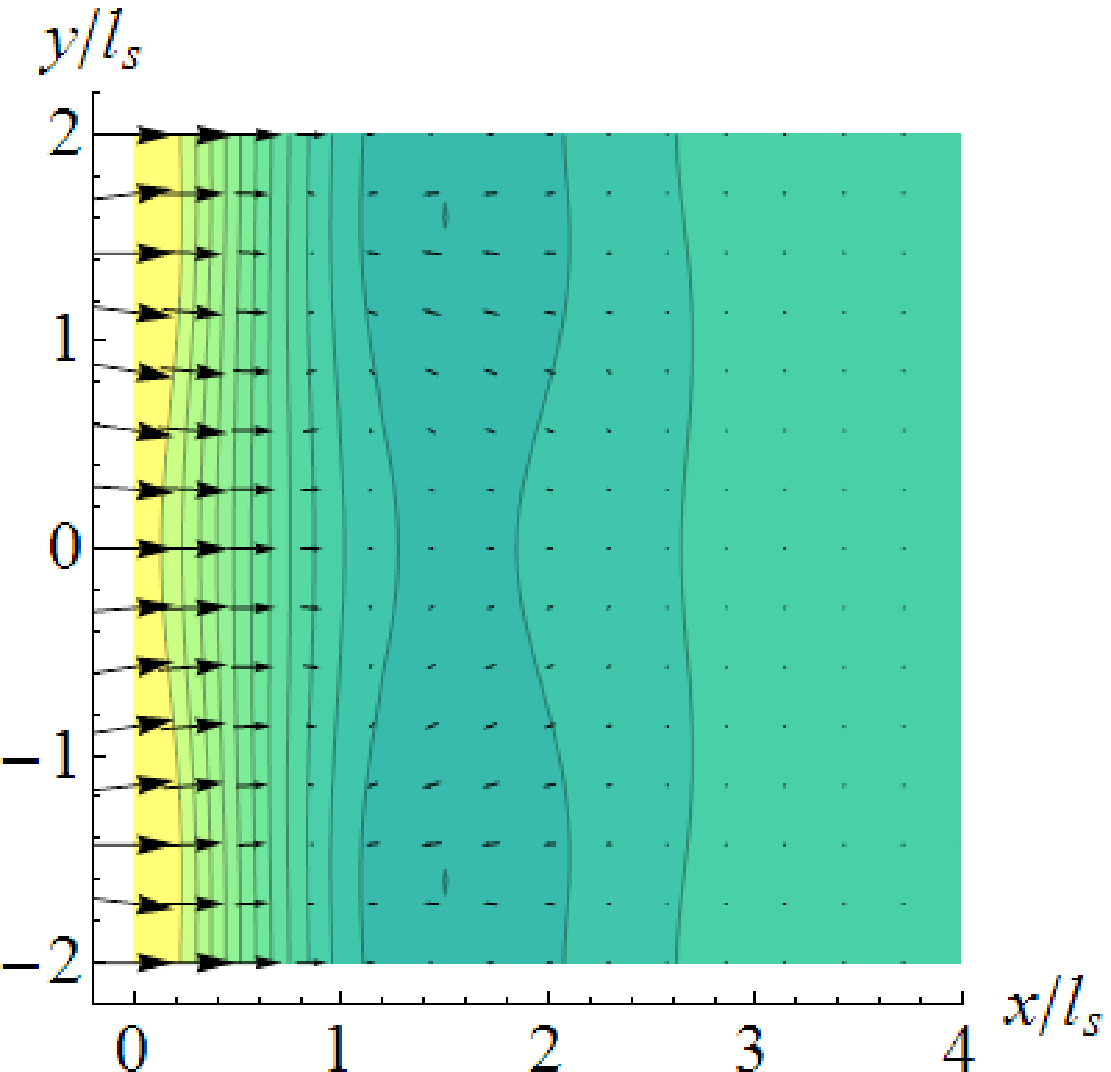}
      \label{fig:int_fx}
	}
      \subfigure[Spin density and spin current polarized along $y$]{
      \includegraphics[scale=0.46]{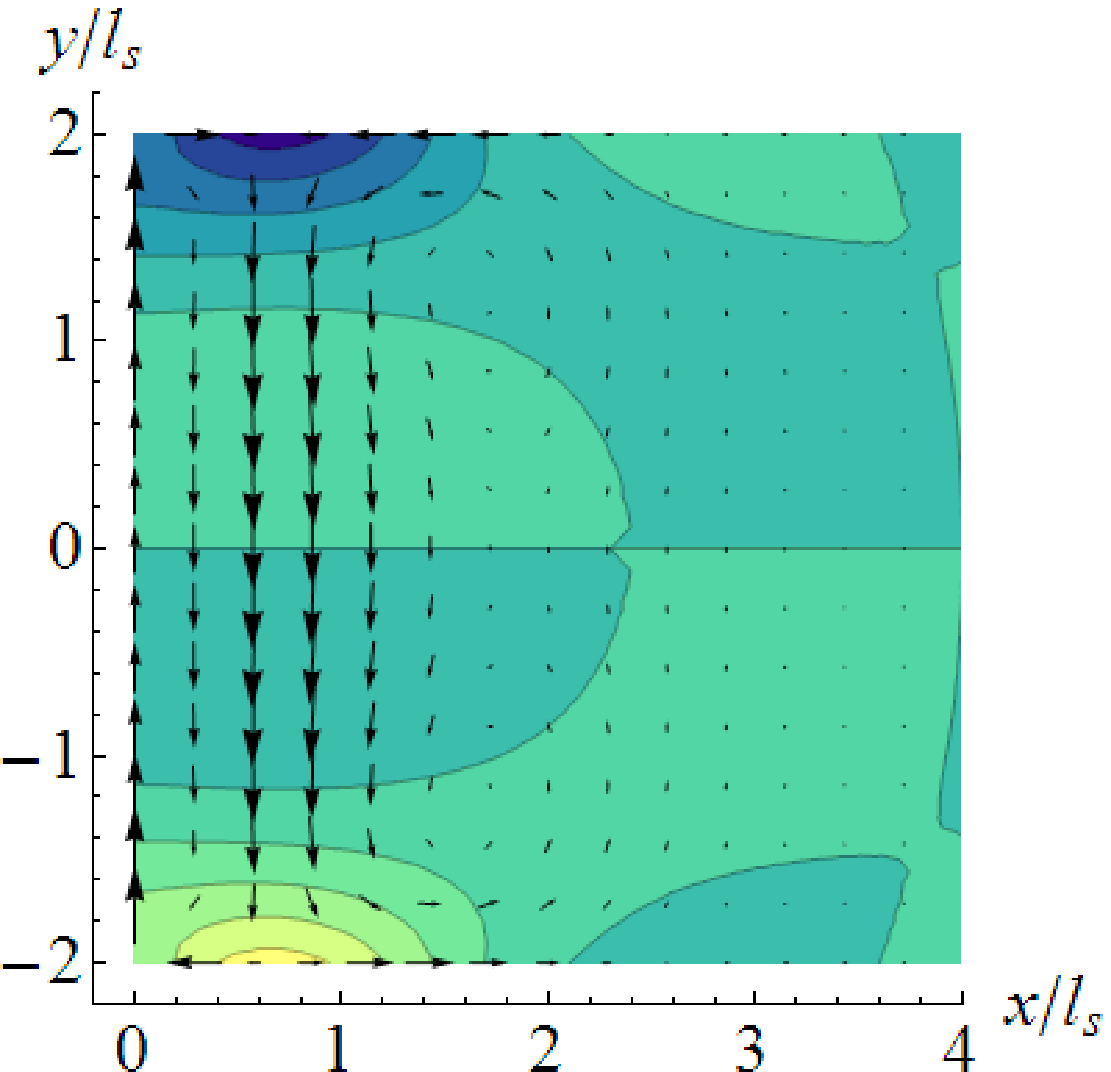}
      \label{fig:int_fy}
	}
      \subfigure[Spin density and spin current polarized along $z$]{
      \includegraphics[scale=0.46]{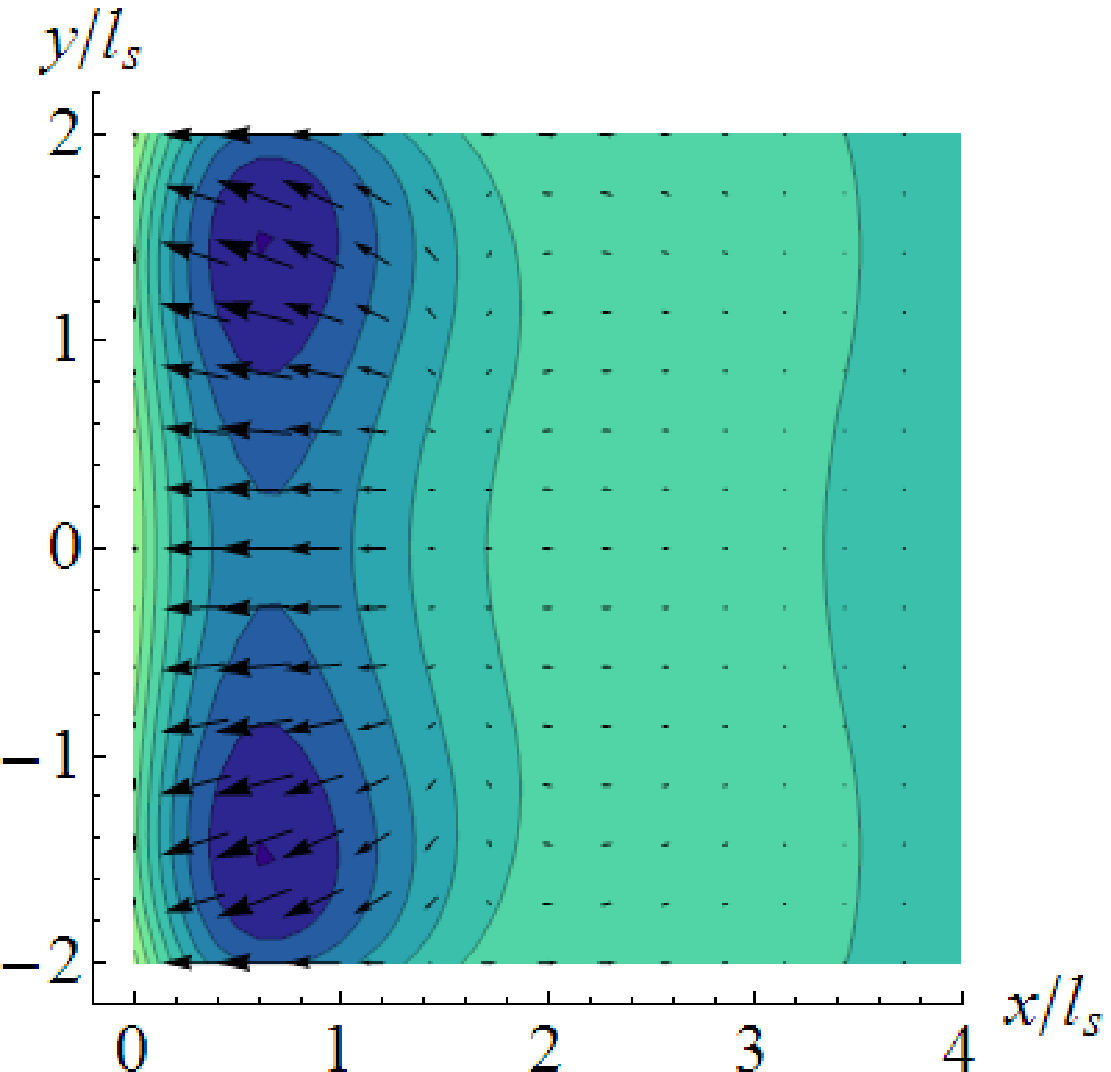}
      \label{fig:int_fz}
	}
      \subfigure{
      \raisebox{34pt}{\setlength\fboxsep{2pt}
\setlength\fboxrule{0.25pt}
      \fbox{\includegraphics[scale=0.46]{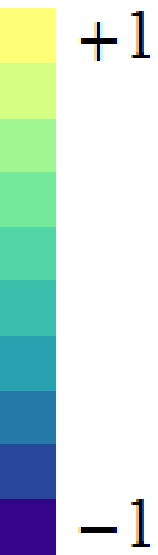}}}
	}
      \caption{(Color online) The intrinsic spin swapping effect in a two-dimensional diffusive metal of width $L = 4 l_\text{s}$ and length $L_x = 16 l_\text{s}$. A primary spin current $j_{x x}^{(0)}$ injected at $x=0$ in \subref{fig:int_fx} induces an oscillating transverse spin current $j_{y y}$ in \subref{fig:int_fy} through coupling with the $z$ components of the spins in \subref{fig:int_fz}. The resulting accumulation of $y$ components of the spins at the sample edges in \subref{fig:int_fy} is a signature of the intrinsic spin swapping effect. Shown are the spin densities and spin current densities according to Eqs.\ \eqref{eq:intrinsic_spin_diffusion} and \eqref{eq:intrinsic_spin_current} on a relative scale for each plot. Note that all quantities are of the same order of magnitude.}
    \label{fig:int_dens}
    \end{figure}
    
    \end{widetext}

%------------------------------------------------------------
% Intrinsic spin swapping
%------------------------------------------------------------

    \section{Intrinsic spin swapping}
    \label{sec:intrinsic}

    We now elucidate the nature of the intrinsic spin swapping effect. The Hamiltonian of a two-dimensional metal with intrinsic spin-orbit coupling reads as
    % -------------------------------------------------------
    % Hamiltonian Intrinsic
    \begin{equation}
    \label{eq:intrinsic_hamiltonian}
      \uppercase{\mathcal{H}}(\vect{k})
      =
      \frac{\hbar^2 k^2}{2m^*}
      +
      \vect{\sigma} \cdot \vect{h}_{\vect{k}}
      +
      v(\vect{k}),
    \end{equation}
    % - - - - - - - - - - - - - - - - - - - - - - - - - - - -
    where $m^*$ is the effective electron mass, $\vect{k}$ is the electron wave vector, and $v(\vect{k})$ is the Fourier transformed scattering potential. We assume Rashba spin-orbit coupling,\cite{Bychkov1984}
    % -------------------------------------------------------
    % Rashba
    \begin{equation}
    \label{eq:rashba_soi}
      \vect{h}_{\vect{k}}
      =
      \big( \alpha k_y, - \alpha k_x, 0 \big)^\text{T}
      ,
    \end{equation}
    % - - - - - - - - - - - - - - - - - - - - - - - - - - - -
    where $\alpha$ defines the spin-orbit coupling strength. When $\alpha$ is sufficiently small, such that the spin-orbit precession length $l_\text{s} = (\alpha m^*)^{-1}$ is much larger than the elastic mean free path, the spin diffusion equation reads \cite{Tang2005}
    % -------------------------------------------------------
    % Spin diffusion
%       \begin{equation}
%       \label{eq:intrinsic_spin_diffusion}
% 	(\partial_x^2 + \partial_y^2) \vect{f}_{\text{s}} + \matrx{C} \vect{f}_{\text{s}} = \matrx{A}_x \partial_x \vect{f}_{\text{s}} + \matrx{A}_y \partial_y \vect{f}_{\text{s}}
%       \end{equation}
    \begin{subequations}
    \label{eq:intrinsic_spin_diffusion}
      \begin{align}
      \label{eq:intrinsic_spin_diffusion_x}
	\vect{\partial}^{2}_{\vect{\rho}} f_x - \frac{4}{l_\text{s}^2} f_x &= \frac{4}{l_\text{s}} \partial_x f_z
	,\\
      \label{eq:intrinsic_spin_diffusion_y}
	\vect{\partial}^{2}_{\vect{\rho}} f_y - \frac{4}{l_\text{s}^2} f_y &= \frac{4}{l_\text{s}} \partial_y f_z
	,\\
      \label{eq:intrinsic_spin_diffusion_z}
	\vect{\partial}^{2}_{\vect{\rho}} f_z - \frac{8}{l_\text{s}^2} f_z &= - \frac{4}{l_\text{s}} \big(\partial_x f_x + \partial_y f_y\big)
	.
      \end{align}
    \end{subequations}
    % - - - - - - - - - - - - - - - - - - - - - - - - - - - -
   The spin current is given by\cite{Tang2005,Brataas2007}
    % -------------------------------------------------------
    % Intrinsic Spin Current
    \begin{equation}
    \label{eq:intrinsic_spin_current}
      j_{a b}
      =
      - D \partial_a f_b
      +
      \frac{2}{l_\text{s}} D \big(
	\delta_{a b}
	f_z
	-
	\delta_{b z}
	f_a
      \big)
      .
    \end{equation}
    The diffusion equations \eqref{eq:intrinsic_spin_diffusion} for the case of intrinsic spin-orbit coupling are more difficult to solve analytically than for the extrinsic case because the $x$, $y$, and $z$ spin components are coupled. Therefore, we numerically study the spin currents and the accumulations of spins resulting from intrinsic spin swapping in a two-dimensional system. Before presenting the numerical results, we discuss the simple analytical expressions that can be derived for the spin accumulations and spin currents induced by intrinsic spin swapping far away from the lateral edges of a sample. The problem can also be treated analytically for a narrow strip system whose width is small compared to the spin-orbit precession length (see Sec.\ \ref{sec:narrowstrip}).
    
    We first consider a case analogous to that given for extrinsic spin swapping. A spin current $j_{x x}^{(0)} = j_{x x}(x = 0)$ carrying spins polarized along the $x$ direction and directed along $x$ is injected at $x=0$.  Again, we assume that the injected current is homogeneous along $y$ at the injection edge and that the lateral edges of the sample are impenetrable, i.e., $j_{y b}(y = \pm L/2) = 0$ for any spin polarization $b$. The situation is, to some extent, similar to the extrinsic case depicted in Fig.\ \ref{fig:ext_dens}. However, while the swapping effect in this scenario is straightforward for the extrinsic case, it is much more complex and rich for intrinsic spin swapping. As mentioned before, analytical expressions can be found for the spin currents and accumulations far away from the lateral boundaries, at distances much larger than $l_\text{s}$. In this region, the influence of the boundaries is weak, and the expressions approach the limit of a system that is infinite in the $y$ direction. We thus find that a transverse spin current $j_{y y}$ flowing along the $y$ direction carrying spins polarized along $y$ is induced,
    % -------------------------------------------------------
    % Induced transverse Spin Currents Intrinsic
    \begin{equation}
    \label{eq:intrinsic_spin_current_y}
    \begin{split}
      &\frac{j_{y y}(x)}{j_{x x}^{(0)}} =
      \\
      &\e{-k_r x/l_\text{s}}
      \Big[
      \big(\! \sqrt{2} - 1 \big) \cos \big(k_i x / l_\text{s} \big)
      - \frac{3 + \! \sqrt{2}}{\sqrt{7}} \, \sin \big(k_i x / l_\text{s} \big)
      \Big]
      ,
    \end{split}
    \end{equation}
    % - - - - - - - - - - - - - - - - - - - - - - - - - - - -
    where $k_{r/i} = \! \sqrt{2 \! \sqrt{2} \mp 1}$. This is the intrinsic spin swapping effect. The induced spin current reaches its maximum, $|j_{y y}(x_\text{max})|/j_{x x}^{(0)} \approx 61\%$, within one spin-orbit precession length from the injection edge at $x=0$. The injected spin current itself decays away from the spin current source at $x=0$,
    % -------------------------------------------------------
    % Induced transverse Spin Currents Intrinsic
    \begin{equation}
    \label{eq:intrinsic_spin_current_x}
      \frac{j_{x x}(x)}{j_{x x}^{(0)}} =
      \e{-k_r x/l_\text{s}}
      \Big[
      \cos \big(k_i x / l_\text{s} \big)
      + \frac{k_r^2}{\sqrt{7}} \, \sin \big(k_i x / l_\text{s} \big)
      \Big]
      .
    \end{equation}
    % - - - - - - - - - - - - - - - - - - - - - - - - - - - -
    
    \begin{widetext}

    \begin{figure}[h!bt]
    \centering
    \subfigure[Spin density and spin current polarized along $x$]{
      \includegraphics[scale=0.46]{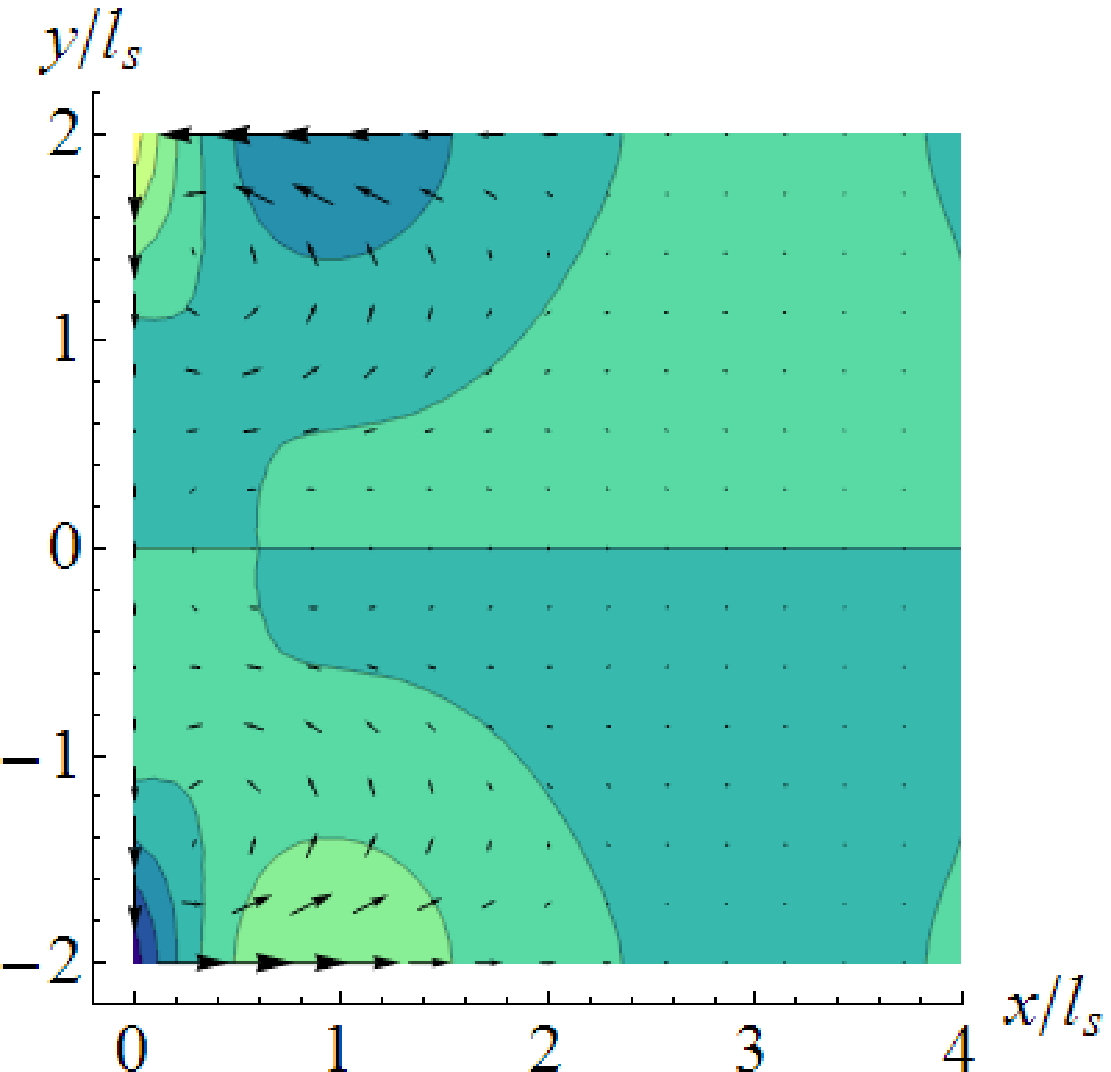}
      \label{fig:int_fx2}
	}
      \subfigure[Spin density and spin current polarized along $y$]{
      \includegraphics[scale=0.46]{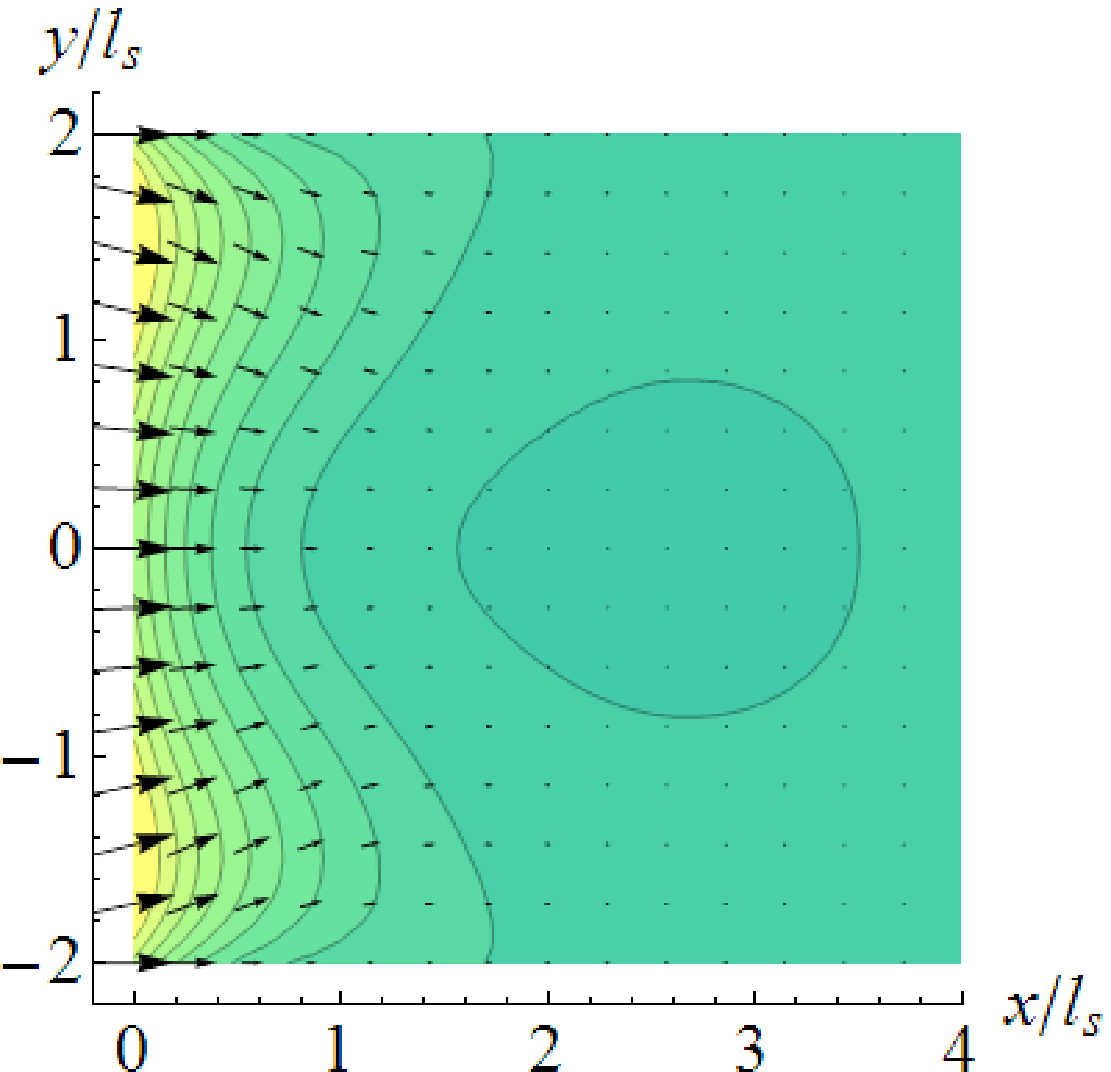}
      \label{fig:int_fy2}
	}
      \subfigure[Spin density and spin current polarized along $z$]{
      \includegraphics[scale=0.46]{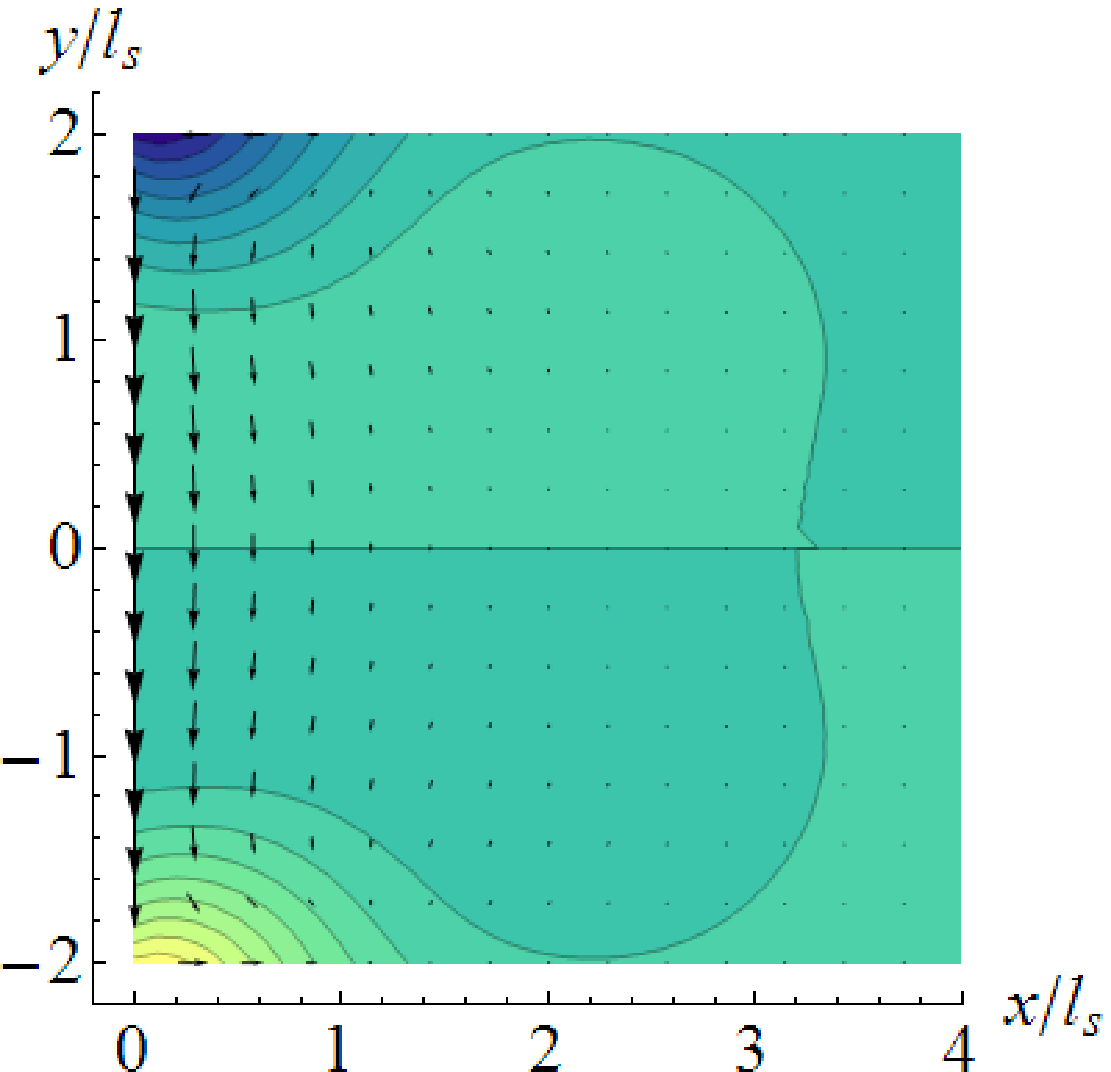}
      \label{fig:int_fz2}
	}
      \subfigure{
      \raisebox{34pt}{\setlength\fboxsep{2pt}
\setlength\fboxrule{0.25pt}
      \fbox{\includegraphics[scale=0.46]{legendv}}}
	}
      \caption{(Color online) The intrinsic spin swapping effect in a two-dimensional diffusive metal of width $L = 4 l_\text{s}$ and length $L_x = 16 l_\text{s}$. A primary spin current $j_{x y}^{(0)}$ injected at $x=0$ in \subref{fig:int_fy2} induces a transverse spin current $j_{y z}$ in \subref{fig:int_fz2}. In turn, this leads to an accumulation of $z$ spins at the sample edges, which is a signature of the intrinsic spin swapping effect. In \subref{fig:int_fx2}, an oscillating spin current polarized along $x$ is only generated close to the lateral edges of the system. Shown are the spin densities and spin current densities according to Eqs.\ \eqref{eq:intrinsic_spin_diffusion} and \eqref{eq:intrinsic_spin_current} on a relative scale for each plot. Note that all quantities are of the same order of magnitude.}
    \label{fig:int_dens2}
    \end{figure}

    \end{widetext}    
    
    While extrinsic spin swapping in general directly couples $x$-polarized and $y$-polarized spins, in intrinsic spin swapping the conversion between $x$-polarized and $y$-polarized spin currents occurs via spins polarized along $z$ as can be seen from Eqs.\ \eqref{eq:intrinsic_spin_diffusion} and \eqref{eq:intrinsic_spin_current}. In addition, spin currents and spin accumulations oscillate as a function of the distance from the injection edge. The situation is depicted in Fig.\ \ref{fig:int_dens} for a system with width $L = 4 l_\text{s}$ and length $L_x = 16 l_\text{s}$. In Fig.\ \ref{fig:int_fx}, we see that the spin current carrying spins polarized along $x$, which is given by Eq.\ \eqref{eq:intrinsic_spin_current_x} in the bulk, as well as the spin accumulation decay away from the spin current source at $x=0$. The $x$ components of the spins are converted to $z$ components, as shown in Fig.\ \ref{fig:int_fz}, which in turn gives rise to a swapped transverse spin current $j_{y y}$, shown in Fig.\ \ref{fig:int_fy}, that is polarized along $y$. In the bulk, this current is given by Eq.\ \eqref{eq:intrinsic_spin_current_y}. We also see that this swapped spin current causes an oscillating spin accumulation at the lateral edges, which is a signature of the intrinsic spin swapping effect that may be probed experimentally (see Sec.\ \ref{sec:narrowstrip} for an explicit expression of this spin swapping induced spin accumulation in a narrow strip system).

    Next, we turn to the case in which a homogeneous spin current $j_{x y}^{(0)} = j_{x y}(x = 0)$ carrying spins polarized along $y$ is injected at $x=0$. To analyze this situation, we first find an analytic expression for the transverse spin current induced through spin swapping far from the lateral edges of the system. We find that the primary spin current $j_{x y}$ is directly transformed into a transverse spin current,
    % -------------------------------------------------------
    % Induced transverse Spin Currents Intrinsic 2
    \begin{equation}
    \label{eq:intrinsic_spin_current_z}
      \frac{j_{y z}(x)}{j_{x y}^{(0)}}
      =
      - \e{-2 x/l_\text{s}}
      =
      - \frac{j_{x y}(x)}{j_{x y}^{(0)}}
      ,
    \end{equation}
    % - - - - - - - - - - - - - - - - - - - - - - - - - - - -
    that gives rise to an accumulation of $z$ spins at the lateral edges of the sample (again, refer to Sec.\ \ref{sec:narrowstrip} for an explicit expression for the induced spin accumulation in a narrow strip system). In contrast to the case of extrinsic spin swapping, Eqs.\ \eqref{eq:intrinsic_spin_diffusion} and \eqref{eq:intrinsic_spin_current} provide a direct coupling between the $y$ and $z$ spins, with the resulting spin current having polarization along $z$ (rather than $x$): again, the resulting current is of the same order as the primary spin current. It is only near the lateral boundaries that spin currents polarized along $x$ are generated as well. This spin current leads to an oscillating spin accumulation at the sample edges. This situation is depicted in Fig.\ \ref{fig:int_dens2}.
    
    In both scenarios of injected spin currents discussed above, intrinsic spin swapping is a much stronger effect than extrinsic spin swapping.

%------------------------------------------------------------
% Narrow Strip
%------------------------------------------------------------

     \section{Intrinsic spin swapping in a narrow strip}
     \label{sec:narrowstrip}

     In this section, we will consider the special case of a strip whose width $L$ is much less than $l_\text{s}$. This case is interesting because, in such a system, a long-range spin swapping effect can be realized, such that the spin-swapped accumulation can extend far along the strip, over a length much greater than $l_\text{s}$. This long-range behavior is closely related to the increase of the D'yakonov-Perel spin relaxation time in narrow strips\cite{Malsh2000}.
     
     Due to the small parameter $L/l_\text{s}$, the spin-swapping problem can be treated analytically. Following Ref.\ \onlinecite{Malsh2000}, we introduce new spin density variables,
     \begin{equation}
     \label{eq:narrow_psi}
       \psi_{\pm 1} = \frac{1}{\sqrt{2}}\big(\pm f_x - \im f_y\big),\quad \psi_{0} = f_z .
     \end{equation}
     In terms of these variables, Eq.\ \eqref{eq:intrinsic_spin_diffusion} can be transformed into
     \begin{equation}
     \label{eq:narrow_psi_diffusion}
       \Big(\im \partial_x + \frac{2}{l_\text{s}} J_y\Big)^2 \vect{\psi} + \Big(\im \partial_y - \frac{2}{l_\text{s}} J_x \Big)^2 \vect{\psi} = 0,
     \end{equation}
     where $\vect{\psi}$ is a 3-vector $(\psi_1,\psi_0,\psi_{-1})^\text{T}$ and $J_i$, $i \in \{x,y,z\}$, are the corresponding 3 $\times$ 3 angular momentum operators for spin $1$. Using Eq.\ \eqref{eq:intrinsic_spin_current}, the boundary conditions can be expressed as
     \begin{subequations}
     \label{eq:narrow_psi_boundary}
     \begin{align}
     \label{eq:narrow_psi_boundary_x}
       &\Big(\im \partial_x + \frac{2}{l_\text{s}} J_y \Big) \vect{\psi} |_{x = 0} = \vect{I},\\
     \label{eq:narrow_psi_boundary_y}
       &\Big(\im \partial_y - \frac{2}{l_\text{s}} J_x \Big) \vect{\psi} |_{y = \pm L/2} = 0,
     \end{align}
     where $\vect{I}$ is determined by the spin current injected at $x = 0$,
     \begin{equation}
     \label{eq:narrow_psi_boundary_I}
       I_{\pm 1} = \frac{\im}{\sqrt{2} D} \Big( \mp j_{x x}^{(0)} + \im j_{x y}^{(0)} \Big),\quad I_0 = - \frac{\im}{D} j_{x z}^{(0)}.
     \end{equation} 
     \end{subequations}      
     The unitary transformation
     \begin{equation}
     \label{eq:narrow_unitary}
       \vect{\psi} = \e{\im J_x(\pi/2 - 2 y/l_\text{s})} \vect{\phi}
     \end{equation}
     further simplifies Eq.\ \eqref{eq:narrow_psi_diffusion} to
     \begin{equation}
     \label{eq:narrow_phi_diffusion}
       \Big(\im \partial_x + \frac{2}{l_\text{s}} J_y(y) \Big)^2 \vect{\phi} - \partial^2_y \vect{\phi} = 0,
     \end{equation}
     where $J_y(y) = \e{-\im J_x(\pi/2 - 2 y/l_\text{s})} J_y \e{\im J_x(\pi/2 - 2 y/l_\text{s})}$, and the boundary conditions at the lateral edges of the system then read
     \begin{equation}
     \label{eq:narrow_phi_boundary_y}
       \partial_y \vect{\phi} |_{y = \pm L/2} = 0.
     \end{equation}
     The transformed differential equations \eqref{eq:narrow_phi_diffusion} and the boundary conditions \eqref{eq:narrow_phi_boundary_y} are exact equivalent representations of the original problem.
     
     For the case of a narrow strip, $L \ll l_\text{s}$, one can expand $J_y(y)$ up to second order in $y/l_\text{s}$ to obtain $J_y(y) = J_z + 2 (y/l_\text{s}) J_y - 2 (y/l_\text{s})^2 J_z$ and consider the last two terms in this expression as a perturbation. Due to Eq.\ \eqref{eq:narrow_phi_boundary_y}, the solution of Eq.\ \eqref{eq:narrow_phi_diffusion} can be represented as a Fourier expansion in $\sin \big( (2 n + 1) \pi y / L \big)$ and  $\cos \big( 2 n \pi y / L \big)$, where $n$ is an integer. Further analysis reveals that only a term uniform in $y$ is relevant because the other Fourier components decay very quickly along the $x$ direction. The equation for $\bar{\vect{\phi}}$, that is, $\vect{\phi}$ averaged over $-L/2 \leq y \leq L/2$, can then be derived from Eq.\ \eqref{eq:narrow_phi_diffusion} as\cite{Malsh2000}
     \begin{equation}
     \label{eq:narrow_barphi_diffusion}
       \Big(\im \partial_x + \frac{2}{l_\text{s}} J_z \Big)^2 \bar{\vect{\phi}} + \frac{\Gamma}{l_\text{s}^2} \Big(2 - J_z^2 \Big) \bar{\vect{\phi}} = 0,
     \end{equation}
     where $\Gamma = 2 L^2 / 3 l_\text{s}^2$. The general solution of this equation that converges for $x \rightarrow \infty$ has the form $\bar{\phi}_{\pm 1} = A_{\pm 1} \e{\pm 2 \, \im x/l_\text{s}} \e{-\sqrt{\Gamma} x/l_\text{s}}$ and $\bar{\phi}_{0} = A_0 \e{-\sqrt{2 \Gamma } x/l_\text{s}}$. The coefficients $A$ can be found from the boundary condition \eqref{eq:narrow_psi_boundary_x}.
     
     If we consider a case analogous to that presented for extrinsic spin swapping in Sec.\ \ref{sec:extrinsic}, where a spin current $j_{x x}^{(0)}$ carrying spins polarized along $x$ is injected, we find $I_{\pm 1} = \mp \frac{\im}{\sqrt{2} D} j_{x x}^{(0)}$ and $I_0 = 0$. Applying the unitary operator \eqref{eq:narrow_unitary} to this boundary condition we obtain in the leading approximation
     \begin{equation}
     \label{eq:narrow_bcphi}
       \bar{\phi}_{\pm 1}(x = 0) = \pm \frac{j_{xx}^{(0)} l_\text{s}}{\sqrt{2 \Gamma} D}, \quad \bar{\phi}_{0}(x = 0) = 0.
     \end{equation}
     From this it follows that
     \begin{equation}
     \label{eq:narrow_phix}
       \bar{\phi}_{\pm 1} = \pm \frac{j_{xx}^{(0)} l_\text{s}}{\sqrt{2 \Gamma} D} \e{\pm 2 \, \im x/l_\text{s} -\sqrt{\Gamma} x/l_\text{s}}, \quad \bar{\phi}_{0} = 0.
     \end{equation}
     Using Eqs.\ \eqref{eq:narrow_unitary} and \eqref{eq:narrow_psi}, we finally obtain the spin densities
     \begin{subequations}
     \label{eq:narrow_spin_density}
     \begin{align}
      \label{eq:narrow_spin_density_x}
	f_x &= \frac{j_{xx}^{(0)} l_\text{s}}{\sqrt{\Gamma} D} \e{-\sqrt{\Gamma} x/l_\text{s}} \, \cos \big(2 x/l_\text{s} \big)
	,\\
      \label{eq:narrow_spin_density_y}
	f_y &= - 2 \frac{j_{xx}^{(0)} l_\text{s}}{\sqrt{\Gamma} D} \frac{y}{l_\text{s}} \e{-\sqrt{\Gamma} x/l_\text{s}} \, \sin \big(2 x/l_\text{s} \big)
	,\\
      \label{eq:narrow_spin_density_z}
	f_z &= - \frac{j_{xx}^{(0)} l_\text{s}}{\sqrt{\Gamma} D} \e{-\sqrt{\Gamma} x/l_\text{s}} \, \sin \big(2 x/l_\text{s} \big)
	,
     \end{align}
     \end{subequations}
     to first order in $y/l_\text{s}$. The accumulation of $y$ spins at the lateral edges of the narrow strip caused by the intrinsic spin swapping effect reads
     \begin{equation}
     \label{eq:narrow_fy}
       f_y(y = \pm L/2) = \mp \sqrt{\frac{3}{2}} \frac{j_{xx}^{(0)} l_\text{s}}{D} \e{-\sqrt{\Gamma} x/l_\text{s}} \, \sin \big(2 x/l_\text{s} \big).
     \end{equation}
     Since $\sqrt{\Gamma}/l_\text{s} \ll 1$, the spin accumulation oscillates and slowly decreases along $x$.
     
     Considering the second case treated in Sec.\ \ref{sec:intrinsic}, where a spin current $j_{x y}^{(0)}$ carrying spins polarized along $y$ is injected, a similar calculation yields
     \begin{subequations}
      \label{eq:narrow_spin_density2}
     \begin{align}
      \label{eq:narrow_spin_density2_x}
	f_x &= 0
	,\\
      \label{eq:narrow_spin_density2_y}
	f_y &= \frac{j_{xy}^{(0)} l_\text{s}}{\sqrt{2 \Gamma} D} \e{-\sqrt{2 \Gamma} x/l_\text{s}}
	,\\
      \label{eq:narrow_spin_density2_z}
	f_z &= - 2 \frac{j_{xy}^{(0)} l_\text{s}}{\sqrt{2 \Gamma} D} \frac{y}{l_\text{s}} \e{-\sqrt{2 \Gamma} x/l_\text{s}}
	.
     \end{align}
     \end{subequations}
     Again, the spin densities slowly decay along $x$ but, analogous to the previous discussion, no oscillation takes place.

%------------------------------------------------------------
% Observation of the Spin Swapping Effect
%------------------------------------------------------------

   \section{Experimental observation of spin swapping}
   \label{sec:observation}
   
    In order to observe spin swapping, a primary spin current needs to be injected. This can be achieved in a two terminal setup where a spin current is electrically injected into a two-dimensional diffusive metal from a ferromagnetic electrode.\cite{Lifshits2009} As discussed here, spin swapping then gives rise to spin accumulations at the lateral sample edges that could be detected experimentally, for example, by optical means \cite{Sih2005} or by measuring the interface voltage at weak contacts between the lateral boundaries and ferromagnets \cite{Valenzuela2006, Saitoh2006,Kimura2007}. However, in such a setup, an electric current is present in the system as well and additional spin currents therefore emerge from the coupling of charge and spin via the spin Hall effect. In a two-dimensional system with extrinsic spin-orbit coupling, the spin accumulations resulting from spin swapping at the lateral sample edges are polarized in-plane while those generated by the electric current via the spin Hall effect are polarized out-of-plane.\cite{Dyakonov1971, Dyakonov1971b} This makes it possible to experimentally distinguish the two effects. On the other hand, in a diffusive system with intrinsic Rashba spin-orbit coupling, a uniform electric field gives rise to a uniform in-plane spin polarization via the Edelstein effect (while it does not produce spin currents).\cite{Edelstein1990, Inoue2003} In contrast, the resulting in-plane accumulation of swapped spins generated by a primary spin current with in-plane polarization is opposite at the lateral boundaries, as discussed above. This difference allows to distinguish the intrinsic spin swapping effect and the Edelstein effect in experiment.
    
    Another possibility is the use of a non-local geometry\cite{Valenzuela2006} where the spin swapping effects could be observed in a part of the system where there is no charge current. There, an electric current is injected from a ferromagnetic electrode on top of a diffusive metal towards a second electrode. A tunnel barrier between the electrodes and the metal assures that the current is injected uniformly and it optimizes the polarization of the injected electrons. A pure spin current is thus generated in the system, propagating in opposite direction of the injected charge current and away from the electrodes. This spin current will give rise to spin accumulations at the lateral sample edges through the spin swapping effect that could be detected experimentally.

%------------------------------------------------------------
% Conclusion
%------------------------------------------------------------

   \section{Conclusion}
   \label{sec:conclusion}

    In conclusion, we have demonstrated that there is an intrinsic analog to the extrinsic spin swapping effect in two-dimensional diffusive metals with Rashba spin-orbit coupling. We found that the intrinsic effect is drastically different because it is large for system dimensions exceeding the spin-orbit precession length and gives rise to secondary spin currents and accumulations that are of the same order of magnitude as the injected primary spin currents while leading to a long-range propagation of spin polarizations in narrow strip systems. In contrast, the extrinsic spin swapping effect is proportional to the spin-orbit coupling strength for any system size and is therefore small. Moreover, intrinsic spin swapping is more complex and richer than its extrinsic counterpart, resulting in a non-trivial dependence on the relative orientation of the injected spin flow and the spin polarization.

     We derived explicit expressions for the transverse spin currents in the bulk and numerically computed the resulting spin accumulations at the lateral boundaries. In addition, we derived explicit expressions for the spin accumulations in a narrow strip when $L \ll l_\text{s}$ and found that the exponential decay of spin polarizations along the $x$ direction is greatly reduced in such systems. We further gave a brief discussion on how the spin swapping effect could be observed in experiment.

   \begin{acknowledgments}
    This work was partially supported by the Research Council of Norway.
   \end{acknowledgments}

\nocite{*}
\bibliographystyle{apsrev4-1}
\bibliography{bibliography}

%Merlin.mbs v4.21 2009-07-09.
\begin{thebibliography}{10}%
\makeatletter
\providecommand \@ifxundefined [1]{%
 \ifx #1\undefined \expandafter \@firstoftwo
 \else \expandafter \@secondoftwo
\fi
}%
\providecommand \@ifnum [1]{%
 \ifnum #1\expandafter \@firstoftwo
 \else \expandafter \@secondoftwo
\fi
}%
\providecommand \enquote [1]{``#1''}%
\providecommand \bibnamefont  [1]{#1}%
\providecommand \bibfnamefont [1]{#1}%
\providecommand \citenamefont [1]{#1}%
\providecommand\href[0]{\@sanitize\@href}%
\providecommand\@href[1]{\endgroup\@@startlink{#1}\endgroup\@@href}%
\providecommand\@@href[1]{#1\@@endlink}%
\providecommand \@sanitize [0]{\begingroup\catcode`\&12\catcode`\#12\relax}%
\@ifxundefined \pdfoutput {\@firstoftwo}{%
 \@ifnum{\z@=\pdfoutput}{\@firstoftwo}{\@secondoftwo}%
}{%
 \providecommand\@@startlink[1]{\leavevmode\special{html:<a href="#1">}}%
 \providecommand\@@endlink[0]{\special{html:</a>}}%
}{%
 \providecommand\@@startlink[1]{%
  \leavevmode
  \pdfstartlink
   attr{/Border[0 0 1 ]/H/I/C[0 1 1]}%
   user{/Subtype/Link/A<</Type/Action/S/URI/URI(#1)>>}%
  \relax
 }%
 \providecommand\@@endlink[0]{\pdfendlink}%
}%
\providecommand \url  [0]{\begingroup\@sanitize \@url }%
\providecommand \@url [1]{\endgroup\@href {#1}{\urlprefix}}%
\providecommand \urlprefix [0]{URL }%
\providecommand \Eprint[0]{\href }%
\@ifxundefined \urlstyle {%
  \providecommand \doi [1]{doi:\discretionary{}{}{}#1}%
}{%
  \providecommand \doi [0]{doi:\discretionary{}{}{}\begingroup
  \urlstyle{rm}\Url }%
}%
\providecommand \doibase [0]{http://dx.doi.org/}%
\providecommand \Doi[1]{\href{\doibase#1}}%
\providecommand \bibAnnote [3]{%
  \BibitemShut{#1}%
  \begin{quotation}\noindent
    \textsc{Key:}\ #2\\\textsc{Annotation:}\ #3%
  \end{quotation}%
}%
\providecommand \bibAnnoteFile [2]{%
  \IfFileExists{#2}{\bibAnnote {#1} {#2} {\input{#2}}}{}%
}%
\providecommand \typeout [0]{\immediate \write \m@ne }%
\providecommand \selectlanguage [0]{\@gobble}%
\providecommand \bibinfo [0]{\@secondoftwo}%
\providecommand \bibfield [0]{\@secondoftwo}%
\providecommand \translation [1]{[#1]}%
\providecommand \BibitemOpen[0]{}%
\providecommand \bibitemStop [0]{}%
\providecommand \bibitemNoStop [0]{.\EOS\space}%
\providecommand \EOS [0]{\spacefactor3000\relax}%
\providecommand \BibitemShut [1]{\csname bibitem#1\endcsname}%
%</preamble>
\bibitem{Karplus1954}%
  \BibitemOpen
  \bibfield{author}{%
  \bibinfo {author} {\bibfnamefont{R.}~\bibnamefont{Karplus}}\ and\ \bibinfo
  {author} {\bibfnamefont{J.~M.}\ \bibnamefont{Luttinger}},\ }%
  \bibfield{journal}{%
  \Doi{10.1103/PhysRev.95.1154}{\bibinfo {journal} {Phys. Rev.}}\ }%
  \textbf{\bibinfo {volume} {95}},\ \bibinfo {pages} {1154} (\bibinfo {year}
  {1954})%
  \bibAnnoteFile{NoStop}{Karplus1954}%
\bibitem{Dyakonov1971b}%
  \BibitemOpen
  \bibfield{author}{%
  \bibinfo {author} {\bibfnamefont{M.~I.}\ \bibnamefont{Dyakonov}}\ and\
  \bibinfo {author} {\bibfnamefont{V.~I.}\ \bibnamefont{Perel}},\ }%
  \bibfield{journal}{%
  \bibinfo {journal} {Sov. Phys. JETP Lett.}\ }%
  \textbf{\bibinfo {volume} {13}},\ \bibinfo {pages} {467} (\bibinfo {year}
  {1971})%
  \bibAnnoteFile{NoStop}{Dyakonov1971b}%
\bibitem{Dyakonov1971}%
  \BibitemOpen
  \bibfield{author}{%
  \bibinfo {author} {\bibfnamefont{M.~I.}\ \bibnamefont{Dyakonov}}\ and\
  \bibinfo {author} {\bibfnamefont{V.~I.}\ \bibnamefont{Perel}},\ }%
  \bibfield{journal}{%
  \Doi{DOI: 10.1016/0375-9601(71)90196-4}{\bibinfo {journal} {Phys. Lett. A}}\
  }%
  \textbf{\bibinfo {volume} {35}},\ \bibinfo {pages} {459} (\bibinfo {year}
  {1971})%
  \bibAnnoteFile{NoStop}{Dyakonov1971}%
\bibitem{Hirsch1999}%
  \BibitemOpen
  \bibfield{author}{%
  \bibinfo {author} {\bibfnamefont{J.~E.}\ \bibnamefont{Hirsch}},\ }%
  \bibfield{journal}{%
  \Doi{10.1103/PhysRevLett.83.1834}{\bibinfo {journal} {Phys. Rev. Lett.}}\ }%
  \textbf{\bibinfo {volume} {83}},\ \bibinfo {pages} {1834} (\bibinfo {year}
  {1999})%
  \bibAnnoteFile{NoStop}{Hirsch1999}%
\bibitem{Zhang2000}%
  \BibitemOpen
  \bibfield{author}{%
  \bibinfo {author} {\bibfnamefont{S.}~\bibnamefont{Zhang}},\ }%
  \bibfield{journal}{%
  \Doi{10.1103/PhysRevLett.85.393}{\bibinfo {journal} {Phys. Rev. Lett.}}\ }%
  \textbf{\bibinfo {volume} {85}},\ \bibinfo {pages} {393} (\bibinfo {year}
  {2000})%
  \bibAnnoteFile{NoStop}{Zhang2000}%
\bibitem{Murakami2003}%
  \BibitemOpen
  \bibfield{author}{%
  \bibinfo {author} {\bibfnamefont{S.}~\bibnamefont{Murakami}}, \bibinfo
  {author} {\bibfnamefont{N.}~\bibnamefont{Nagaosa}},\ and\ \bibinfo {author}
  {\bibfnamefont{S.-C.}\ \bibnamefont{Zhang}},\ }%
  \bibfield{journal}{%
  \Doi{10.1126/science.1087128}{\bibinfo {journal} {Science}}\ }%
  \textbf{\bibinfo {volume} {301}},\ \bibinfo {pages} {1348} (\bibinfo {year}
  {2003})%
  \bibAnnoteFile{NoStop}{Murakami2003}%
\bibitem{Sinova2004}%
  \BibitemOpen
  \bibfield{author}{%
  \bibinfo {author} {\bibfnamefont{J.}~\bibnamefont{Sinova}}, \bibinfo {author}
  {\bibfnamefont{D.}~\bibnamefont{Culcer}}, \bibinfo {author}
  {\bibfnamefont{Q.}~\bibnamefont{Niu}}, \bibinfo {author}
  {\bibfnamefont{N.~A.}\ \bibnamefont{Sinitsyn}}, \bibinfo {author}
  {\bibfnamefont{T.}~\bibnamefont{Jungwirth}},\ and\ \bibinfo {author}
  {\bibfnamefont{A.~H.}\ \bibnamefont{MacDonald}},\ }%
  \bibfield{journal}{%
  \Doi{10.1103/PhysRevLett.92.126603}{\bibinfo {journal} {Phys. Rev. Lett.}}\
  }%
  \textbf{\bibinfo {volume} {92}},\ \bibinfo {pages} {126603} (\bibinfo {year}
  {2004})%
  \bibAnnoteFile{NoStop}{Sinova2004}%
\bibitem{Engel2007}%
  \BibitemOpen
  \bibfield{author}{%
  \bibinfo {author} {\bibfnamefont{H.-A.}\ \bibnamefont{Engel}}, \bibinfo
  {author} {\bibfnamefont{E.~I.}\ \bibnamefont{Rashba}},\ and\ \bibinfo
  {author} {\bibfnamefont{B.~I.}\ \bibnamefont{Halperin}},\ }%
  in\ \emph{\bibinfo {booktitle} {Handbook of Magnetism and Advanced Magnetic
  Materials}},\ Vol.~\bibinfo {volume} {5},\ \bibinfo {editor} {edited by\
  \bibinfo {editor} {\bibfnamefont{H.}~\bibnamefont{Kronm{\"u}ller}}\ and\
  \bibinfo {editor} {\bibfnamefont{S.}~\bibnamefont{Parkin}}}\ (\bibinfo
  {publisher} {Wiley},\ \bibinfo {address} {Chichester},\ \bibinfo {year}
  {2007})\ pp.\ \bibinfo {pages} {2858--2877}%
  \bibAnnoteFile{NoStop}{Engel2007}%
\bibitem{Kato2004}%
  \BibitemOpen
  \bibfield{author}{%
  \bibinfo {author} {\bibfnamefont{Y.~K.}\ \bibnamefont{Kato}}, \bibinfo
  {author} {\bibfnamefont{R.~C.}\ \bibnamefont{Myers}}, \bibinfo {author}
  {\bibfnamefont{A.~C.}\ \bibnamefont{Gossard}},\ and\ \bibinfo {author}
  {\bibfnamefont{D.~D.}\ \bibnamefont{Awschalom}},\ }%
  \bibfield{journal}{%
  \Doi{10.1126/science.1105514}{\bibinfo {journal} {Science}}\ }%
  \textbf{\bibinfo {volume} {306}},\ \bibinfo {pages} {1910} (\bibinfo {year}
  {2004})%
  \bibAnnoteFile{NoStop}{Kato2004}%
\bibitem{Sih2005}%
  \BibitemOpen
  \bibfield{author}{%
  \bibinfo {author} {\bibfnamefont{V.}~\bibnamefont{Sih}}, \bibinfo {author}
  {\bibfnamefont{R.~C.}\ \bibnamefont{Myers}}, \bibinfo {author}
  {\bibfnamefont{Y.~K.}\ \bibnamefont{Kato}}, \bibinfo {author}
  {\bibfnamefont{W.~H.}\ \bibnamefont{Lau}}, \bibinfo {author}
  {\bibfnamefont{A.~C.}\ \bibnamefont{Gossard}},\ and\ \bibinfo {author}
  {\bibfnamefont{D.~D.}\ \bibnamefont{Awschalom}},\ }%
  \bibfield{journal}{%
  \Doi{10.1038/nphys009}{\bibinfo {journal} {Nat. Phys.}}\ }%
  \textbf{\bibinfo {volume} {1}},\ \bibinfo {pages} {31} (\bibinfo {year}
  {2005})%
  \bibAnnoteFile{NoStop}{Sih2005}%
\bibitem{Wunderlich2005}%
  \BibitemOpen
  \bibfield{author}{%
  \bibinfo {author} {\bibfnamefont{J.}~\bibnamefont{Wunderlich}}, \bibinfo
  {author} {\bibfnamefont{B.}~\bibnamefont{Kaestner}}, \bibinfo {author}
  {\bibfnamefont{J.}~\bibnamefont{Sinova}},\ and\ \bibinfo {author}
  {\bibfnamefont{T.}~\bibnamefont{Jungwirth}},\ }%
  \bibfield{journal}{%
  \Doi{10.1103/PhysRevLett.94.047204}{\bibinfo {journal} {Phys. Rev. Lett.}}\
  }%
  \textbf{\bibinfo {volume} {94}},\ \bibinfo {pages} {047204} (\bibinfo {year}
  {2005})%
  \bibAnnoteFile{NoStop}{Wunderlich2005}%
\bibitem{Saitoh2006}%
  \BibitemOpen
  \bibfield{author}{%
  \bibinfo {author} {\bibfnamefont{E.}~\bibnamefont{Saitoh}}, \bibinfo {author}
  {\bibfnamefont{M.}~\bibnamefont{Ueda}}, \bibinfo {author}
  {\bibfnamefont{H.}~\bibnamefont{Miyajima}},\ and\ \bibinfo {author}
  {\bibfnamefont{G.}~\bibnamefont{Tatara}},\ }%
  \bibfield{journal}{%
  \Doi{10.1063/1.2199473}{\bibinfo {journal} {Appl. Phys. Lett.}}\ }%
  \textbf{\bibinfo {volume} {88}},\ \bibinfo {pages} {182509} (\bibinfo {year}
  {2006})%
  \bibAnnoteFile{NoStop}{Saitoh2006}%
\bibitem{Stern2006}%
  \BibitemOpen
  \bibfield{author}{%
  \bibinfo {author} {\bibfnamefont{N.~P.}\ \bibnamefont{Stern}}, \bibinfo
  {author} {\bibfnamefont{S.}~\bibnamefont{Ghosh}}, \bibinfo {author}
  {\bibfnamefont{G.}~\bibnamefont{Xiang}}, \bibinfo {author}
  {\bibfnamefont{M.}~\bibnamefont{Zhu}}, \bibinfo {author}
  {\bibfnamefont{N.}~\bibnamefont{Samarth}},\ and\ \bibinfo {author}
  {\bibfnamefont{D.~D.}\ \bibnamefont{Awschalom}},\ }%
  \bibfield{journal}{%
  \Doi{10.1103/PhysRevLett.97.126603}{\bibinfo {journal} {Phys. Rev. Lett.}}\
  }%
  \textbf{\bibinfo {volume} {97}},\ \bibinfo {pages} {126603} (\bibinfo {year}
  {2006})%
  \bibAnnoteFile{NoStop}{Stern2006}%
\bibitem{Valenzuela2006}%
  \BibitemOpen
  \bibfield{author}{%
  \bibinfo {author} {\bibfnamefont{S.}~\bibnamefont{Valenzuela}}\ and\ \bibinfo
  {author} {\bibfnamefont{M.}~\bibnamefont{Tinkham}},\ }%
  \bibfield{journal}{%
  \Doi{10.1038/nature04937}{\bibinfo {journal} {Nature}}\ }%
  \textbf{\bibinfo {volume} {442}},\ \bibinfo {pages} {176} (\bibinfo {year}
  {2006})%
  \bibAnnoteFile{NoStop}{Valenzuela2006}%
\bibitem{Kimura2007}%
  \BibitemOpen
  \bibfield{author}{%
  \bibinfo {author} {\bibfnamefont{T.}~\bibnamefont{Kimura}}, \bibinfo {author}
  {\bibfnamefont{Y.}~\bibnamefont{Otani}}, \bibinfo {author}
  {\bibfnamefont{T.}~\bibnamefont{Sato}}, \bibinfo {author}
  {\bibfnamefont{S.}~\bibnamefont{Takahashi}},\ and\ \bibinfo {author}
  {\bibfnamefont{S.}~\bibnamefont{Maekawa}},\ }%
  \bibfield{journal}{%
  \Doi{10.1103/PhysRevLett.98.156601}{\bibinfo {journal} {Phys. Rev. Lett.}}\
  }%
  \textbf{\bibinfo {volume} {98}},\ \bibinfo {pages} {156601} (\bibinfo {year}
  {2007})%
  \bibAnnoteFile{NoStop}{Kimura2007}%
\bibitem{Lifshits2009}%
  \BibitemOpen
  \bibfield{author}{%
  \bibinfo {author} {\bibfnamefont{M.~B.}\ \bibnamefont{Lifshits}}\ and\
  \bibinfo {author} {\bibfnamefont{M.~I.}\ \bibnamefont{Dyakonov}},\ }%
  \bibfield{journal}{%
  \Doi{10.1103/PhysRevLett.103.186601}{\bibinfo {journal} {Phys. Rev. Lett.}}\
  }%
  \textbf{\bibinfo {volume} {103}},\ \bibinfo {eid} {186601} (\bibinfo {year}
  {2009})%
  \bibAnnoteFile{NoStop}{Lifshits2009}%
\bibitem{Note1}%
  \BibitemOpen
  \bibinfo {note} {The terms responsible for extrinsic spin swapping were
  already indicated in Refs.\ \protect \onlinecite
  {Dyakonov1971b,Dyakonov1971}. However, their physical origin was not
  understood at the time\protect \cite {Lifshits2009}}%
  \bibAnnoteFile{NoStop}{Note1}%
\bibitem{Malsh2000}%
  \BibitemOpen
  \bibfield{author}{%
  \bibinfo {author} {\bibfnamefont{A.~G.}\ \bibnamefont{Mal'shukov}}\ and\
  \bibinfo {author} {\bibfnamefont{K.~A.}\ \bibnamefont{Chao}},\ }%
  \bibfield{journal}{%
  \Doi{10.1103/PhysRevB.61.R2413}{\bibinfo {journal} {Phys. Rev. B: Condens.
  Matter}}\ }%
  \textbf{\bibinfo {volume} {61}},\ \bibinfo {pages} {R2413} (\bibinfo {year}
  {2000})%
  \bibAnnoteFile{NoStop}{Malsh2000}%
\bibitem{Bychkov1984}%
  \BibitemOpen
  \bibfield{author}{%
  \bibinfo {author} {\bibfnamefont{Y.~A.}\ \bibnamefont{Bychkov}}\ and\
  \bibinfo {author} {\bibfnamefont{E.~I.}\ \bibnamefont{Rashba}},\ }%
  \bibfield{journal}{%
  \Doi{10.1088/0022-3719/17/33/015}{\bibinfo {journal} {J. Phys. C: Solid State
  Phys.}}\ }%
  \textbf{\bibinfo {volume} {17}},\ \bibinfo {pages} {6039} (\bibinfo {year}
  {1984})%
  \bibAnnoteFile{NoStop}{Bychkov1984}%
\bibitem{Tang2005}%
  \BibitemOpen
  \bibfield{author}{%
  \bibinfo {author} {\bibfnamefont{C.~S.}\ \bibnamefont{Tang}}, \bibinfo
  {author} {\bibfnamefont{A.~G.}\ \bibnamefont{Mal'shukov}},\ and\ \bibinfo
  {author} {\bibfnamefont{K.~A.}\ \bibnamefont{Chao}},\ }%
  \bibfield{journal}{%
  \Doi{10.1103/PhysRevB.71.195314}{\bibinfo {journal} {Phys. Rev. B: Condens.
  Matter}}\ }%
  \textbf{\bibinfo {volume} {71}},\ \bibinfo {pages} {195314} (\bibinfo {year}
  {2005})%
  \bibAnnoteFile{NoStop}{Tang2005}%
\bibitem{Brataas2007}%
  \BibitemOpen
  \bibfield{author}{%
  \bibinfo {author} {\bibfnamefont{A.}~\bibnamefont{Brataas}}, \bibinfo
  {author} {\bibfnamefont{A.~G.}\ \bibnamefont{Mal'shukov}},\ and\ \bibinfo
  {author} {\bibfnamefont{Y.}~\bibnamefont{Tserkovnyak}},\ }%
  \bibfield{journal}{%
  \bibinfo {journal} {New J. Phys.}\ }%
  \textbf{\bibinfo {volume} {9}},\ \bibinfo {pages} {345} (\bibinfo {year}
  {2007})%
  \bibAnnoteFile{NoStop}{Brataas2007}%
\bibitem{Edelstein1990}%
  \BibitemOpen
  \bibfield{author}{%
  \bibinfo {author} {\bibfnamefont{V.}~\bibnamefont{Edelstein}},\ }%
  \bibfield{journal}{%
  \Doi{10.1016/0038-1098(90)90963-C}{\bibinfo {journal} {Solid State Commun.}}\
  }%
  \textbf{\bibinfo {volume} {73}},\ \bibinfo {pages} {233} (\bibinfo {year}
  {1990})%
  \bibAnnoteFile{NoStop}{Edelstein1990}%
\bibitem{Inoue2003}%
  \BibitemOpen
  \bibfield{author}{%
  \bibinfo {author} {\bibfnamefont{J.-i.}\ \bibnamefont{Inoue}}, \bibinfo
  {author} {\bibfnamefont{G.~E.~W.}\ \bibnamefont{Bauer}},\ and\ \bibinfo
  {author} {\bibfnamefont{L.~W.}\ \bibnamefont{Molenkamp}},\ }%
  \bibfield{journal}{%
  \Doi{10.1103/PhysRevB.67.033104}{\bibinfo {journal} {Phys. Rev. B: Condens.
  Matter}}\ }%
  \textbf{\bibinfo {volume} {67}},\ \bibinfo {pages} {033104} (\bibinfo {year}
  {2003})%
  \bibAnnoteFile{NoStop}{Inoue2003}%
\bibitem{Rammer2007}%
  \BibitemOpen
  \bibfield{author}{%
  \bibinfo {author} {\bibfnamefont{J.}~\bibnamefont{Rammer}},\ }%
  \emph{\bibinfo {title} {Quantum Field Theory of Non-Equi\-li\-bri\-um
  States}}\ (\bibinfo {publisher} {Cambridge University Press},\ \bibinfo
  {address} {Cambridge},\ \bibinfo {year} {2007})%
  \bibAnnoteFile{NoStop}{Rammer2007}%
\bibitem{Raimondi2009}%
  \BibitemOpen
  \bibfield{author}{%
  \bibinfo {author} {\bibfnamefont{R.}~\bibnamefont{Raimondi}}\ and\ \bibinfo
  {author} {\bibfnamefont{P.}~\bibnamefont{Schwab}},\ }%
  \bibfield{journal}{%
  \Doi{10.1016/j.physe.2009.10.047}{\bibinfo {journal} {Physica E}}\ }%
  \textbf{\bibinfo {volume} {42}},\ \bibinfo {pages} {952 } (\bibinfo {year}
  {2010})%
  \bibAnnoteFile{NoStop}{Raimondi2009}%
\end{thebibliography}%

\end{document}